\documentclass[groupedaddress,superscriptaddress,aps,prl,preprintnumbers,twocolumn,10pt,fleqn,nofootinbib,nobibnote,eqsecnum,floatfix,showpacs,longbibliography]{revtex4-2}

\usepackage{graphics}      
\usepackage{url}           
\usepackage[utf8]{inputenc}

\usepackage{graphicx}
\usepackage{bm,amsmath,amssymb}
\usepackage[mathscr]{eucal}
\usepackage{color}
\usepackage{afterpage}

\newcommand{\der}{\mathrm{d}}

\long\def\comment#1{ }

\newcommand{\eqn}[1]{Eq.~\eqref{#1}}
\newcommand{\beq}{\begin{equation}}
\newcommand{\eeq}{\end{equation}}
\newcommand{\bal}{\begin{align}}
\newcommand{\eal}{\end{align}}

\newcommand{\rmK}{{\rm K}}

\newcommand{\xbj}{x_{_{\rm Bj}}}

\newcommand{\order}[1]{\mathcal{O}{(#1)}}

\newcommand{\rmd}{{\rm d}}

\newcommand{\bk}{\bm{k}}
\newcommand{\bK}{\bm{K}}
\newcommand{\bP}{\bm{P}}

\newcommand{\bx}{\bm{x}}
\newcommand{\by}{\bm{y}}

\newcommand{\bz}{\bm{z}}

\newcommand{\br}{\bm{r}}

\newcommand{\mcal}{\mathcal}

\begin{document}

\title{Probing parton saturation and the gluon dipole via diffractive jet production\\ at the Electron-Ion Collider}

%

\author{E.~Iancu}
\email{edmond.iancu@ipht.fr}
\affiliation{Universit\'{e} Paris-Saclay, CNRS, CEA, Institut de physique th\'{e}orique, F-91191, Gif-sur-Yvette, France}

\author{A.H.~Mueller}
\email{ahm4@columbia.edu}
\affiliation{Department of Physics, Columbia University, New York, NY 10027, USA}

\author{D.N.~Triantafyllopoulos}
\email{trianta@ectstar.eu}
\affiliation{European Centre for Theoretical Studies in Nuclear Physics and Related Areas (ECT*)\\
and Fondazione Bruno Kessler, Strada delle Tabarelle 286, I-38123 Villazzano (TN), Italy}

\date{\today}

\begin{abstract}
We demonstrate that hard dijet production via coherent inelastic diffraction is a promising channel for probing gluon saturation at the Electron-Ion Collider. By {\it inelastic} diffraction we mean a process in which the two hard jets 
--- a quark-antiquark pair generated by the decay of the virtual photon --- 
are accompanied by a softer gluon jet, emitted by the quark or the antiquark.
This  process can be described as the elastic scattering of an effective gluon-gluon dipole. 
The cross section takes a factorised form, between a hard factor and a unintegrated (``Pomeron'')
gluon distribution describing the transverse momentum imbalance between the hard dijets.
The dominant contribution comes from the black disk limit and leads to a dijet imbalance of the order of the
target saturation momentum $Q_s$ evaluated at the rapidity gap.  Integrating out the dijet imbalance, we obtain a collinear factorization where the initial condition for the DGLAP evolution is set by gluon saturation.
    \end{abstract}
    \maketitle

\paragraph{Introduction.}
A class of observables that attracted much attention over the last years refers to the production
of a pair of jets in ``dilute-dense'' collisions (electron-nucleus, proton-nucleus, ultraperipheral proton-nucleus
and nucleus-nucleus)  at high energy \cite{Marquet:2007vb,Albacete:2010pg,Dominguez:2011wm,Metz:2011wb,Dominguez:2011br,Stasto:2011ru,Lappi:2012nh,Iancu:2013dta,Zheng:2014vka,Altinoluk:2015dpi,Hatta:2016dxp,Dumitru:2015gaa,Kotko:2015ura,Marquet:2016cgx,vanHameren:2016ftb,Marquet:2017xwy,Albacete:2018ruq,Dumitru:2018kuw,Mantysaari:2019csc,Salazar:2019ncp,Mantysaari:2019hkq,Boussarie:2021ybe,Kotko:2017oxg,Hagiwara:2017fye,Klein:2019qfb,Hatta:2021jcd,Iancu:2020mos,Caucal:2021ent}.
Through the correlations among the produced jets, these observables have the potential
to probe fine aspects of the gluon distribution in the dense target, like the onset of saturation \cite{Iancu:2003xm,Gelis:2010nm,Kovchegov:2012mbw}, 
or the spatial distribution in the plane transverse to the collision axis \cite{Altinoluk:2015dpi,Hatta:2016dxp}.
It was furthermore observed that the physical content of such observables becomes
more transparent in the ``correlation limit'' where the two measured jets are relatively hard and 
propagate nearly back-to-back in the transverse plane \cite{Dominguez:2011wm,Metz:2011wb,Dominguez:2011br,Dumitru:2015gaa,Kotko:2015ura,Marquet:2016cgx,vanHameren:2016ftb,Albacete:2018ruq,Dumitru:2018kuw}. 
This means that their individual 
transverse momenta $k_{1\perp} =|\bk_1|$ and  $k_{2\perp} =|\bk_2|$ are much larger than their
transverse imbalance:  $k_{1\perp} \simeq  k_{2\perp} \gg K_\perp\equiv |\bk_1+\bk_2|$. 
This limit often permits to factorise the interesting correlations (which refer to the distribution w.r.t.  
$\bK$) from the comparatively hard physics of dijet production.

Specialising to the case of electron-nucleus deep inelastic scattering (DIS) at high energies,
one particular example that serves as a benchmark for our
new results, is the inclusive dijet production. In the correlation limit, the respective cross-section (as computed
in the CGC formalism \cite{Iancu:2003xm,Gelis:2010nm,Kovchegov:2012mbw})
takes a factorised form,  recognised as  the high-energy limit of the TMD factorisation \cite{Dominguez:2011wm}: it is 
the product between a {\it hard factor} describing the photon dissociation into a quark-antiquark ($q\bar q$)
pair (a ``colour dipole'') together with the coupling between this pair and the target gluons,
and a {\it unintegrated gluon distribution}  (UGD), the ``Weiszäcker-Williams gluon TMD'', describing the  transverse 
momentum transfer from the target to the  $q\bar q$ dijet, via inelastic collisions. 

In the presence of gluon saturation, the typical transferred momentum cannot be smaller than  the target
saturation momentum $Q_s(Y)$ at the rapidity scale $Y$ probed by the scattering. The
correlation limit applies when  $k_{1\perp}\,, k_{2\perp} \gg Q_s(Y)$. This in turn requires a relatively hard DIS
process, with $Q^2 \gg Q_s^2(Y)$. However, very large values of $Q^2$ lead to an enhanced radiation
in the final state, which may obscure the physics of
saturation, due to the Sudakov effect \cite{Mueller:2013wwa}: the recoil associated with these emissions 
can dominate the dijet momentum imbalance, 
to the detriment of gluon saturation \cite{Zheng:2014vka}.

\paragraph{Diffractive trijet production in the correlation limit.}
In this Letter, we propose another hard dijet process in DIS at high energy, which is
even more sensitive to gluon saturation than the inclusive dijet production.
 This is the production of a pair of hard jets in the correlation limit, via coherent inelastic diffraction.
 
 ``Diffraction'' refers to a process in which there is a large rapidity gap between the produced jets and the nuclear target, 
 while ``inelastic'' means that the two hard jets --- the $q\bar q$ pair generated by the decay of the virtual photon --- 
 are accompanied by a softer gluon  ($g$)   jet, with transverse momentum $\bk_3$,
 emitted by the quark or the antiquark.
 ``Coherent'' means that the hadronic target (proton or nucleus) does not break in the final state,
so the rapidity gap (denoted as $Y_{\mathbb{P}}$) lies on the target side. 

In such a coherent process, the scattering between the $q\bar q g$ system and
the hadronic target is necessarily elastic. This implies that the transverse momentum
transferred by the target is quite small, $\Delta_\perp\sim 2/R\sim\Lambda$ ($R$ is
the target radius and $\Lambda$ the QCD confinement scale). This is
 negligible compared to the recoil associated with the gluon emission, which therefore
 controls the transverse momentum imbalance $\bK= \bk_1+\bk_2$  between the two hard jets:  $\bK\simeq -\bk_3$. 
The ``correlation limit'' of interest corresponds to
 $k_{1\perp}\simeq k_{2\perp}\gg k_{3\perp}$. This is in fact the {\it typical} trijet 
configuration for sufficiently hard diffraction, $Q^2\gg Q_s^2(Y_{\mathbb{P}})$, as we now explain. 

The quark and the antiquark produced by the decay of a hard virtual photon
have transverse momenta of the order of the virtuality, $ k_{1\perp}\,, k_{2,\perp} \sim Q$, 
whereas the momentum  $k_{3\perp}$ of the gluon jet 
is controlled by the scattering and is typically of order $Q_s(Y_{\mathbb{P}})$.
The last point is specific to diffraction:
the elastic cross-section involves the {\it square} of the forward scattering amplitude, hence it is
more sensitive than the total cross-section to the  ``black disk limit'' where the scattering is strong
and $k_{3\perp}\sim Q_s(Y_{\mathbb{P}})$.
 
 In practice, we propose the experimental measurement of the hard dijets {\it alone}.
 But albeit not directly measured, the  comparatively soft, gluon, jet has a crucial influence on the structure of
 the final state: \texttt{(i)} it controls the momentum imbalance between the two hard jets, and 
  \texttt{(ii)} it opens up the colour space and thus yields a large size 
partonic configuration to probe gluon saturation in the target.

 Diffractive trijet production has also been addressed in the context of $k_T$-factorisation,
 more than  2 decades ago \cite{Bartels:1999tn}.
 However that early study has overlooked the key role of  gluon saturation. 
  The effects of saturation have been included (within the dipole picture) in \cite{Kovchegov:2001ni},
 but the focus there was on diffractive {\it gluon} production only.
 

\paragraph{The kinematics for trijet diffractive production.}
We describe DIS within the dipole picture,
applicable at small Bjorken $\xbj\equiv Q^2/(2 q\cdot P_N)\ll 1$,
with $q^\mu$ and $P^\mu_N$ the 4-momenta of the virtual photon and, respectively,
of a nucleon from the target (assumed to be massless). 
We work in a frame where $q^\mu= (q^+, -Q^2/2q^+,\bm{0} )$ and $P^\mu_N = (0, P^-_N,\bm{0})$, 
in light-cone notations.

We denote the 4-momenta of the produced partons as $k_i^\mu=(k_i^+,k_i^-,\bk_i)$,
with $k_i^-=k_{i\perp}^2/2k_i^+$,
where $i=1,2,3$ refers to the quark, the antiquark, and the gluon, respectively. We shall mostly work with the longitudinal
fractions $\vartheta_i=k_i^+/q^+$, with $\vartheta_1+\vartheta_2+\vartheta_3=1$, and we shall denote
$\xi\equiv\vartheta_3$ for the gluon. We anticipate that the interesting situation is such that $\xi\ll 1$,
whereas $\vartheta_1$ and  $\vartheta_2$ are comparable with each other. 

For the $q\bar q$ pair, we replace $\bm{k}_1 $ and $\bm{k}_2$ with $\bm{P}$ and $\bm{K} $, where
  $\bm{P} = \vartheta_2 \bm{k}_1 - \vartheta_1 \bm{k}_2$ is the relative momentum and
    $\bm{K} = \bm{k}_1+\bm{k}_2$ is the total momentum.
We neglect the transverse momentum transfer from the target,
 $\bk_1+\bk_2+\bk_3=0$, and focus on the correlation limit for the hard dijets:  $P_\perp  \sim Q \gg K_\perp
 = k_{3\perp} \sim Q_s(Y_{\mathbb{P}}) $. 

\begin{figure}[t] \centerline{
\centerline{\includegraphics[width=0.8\columnwidth]{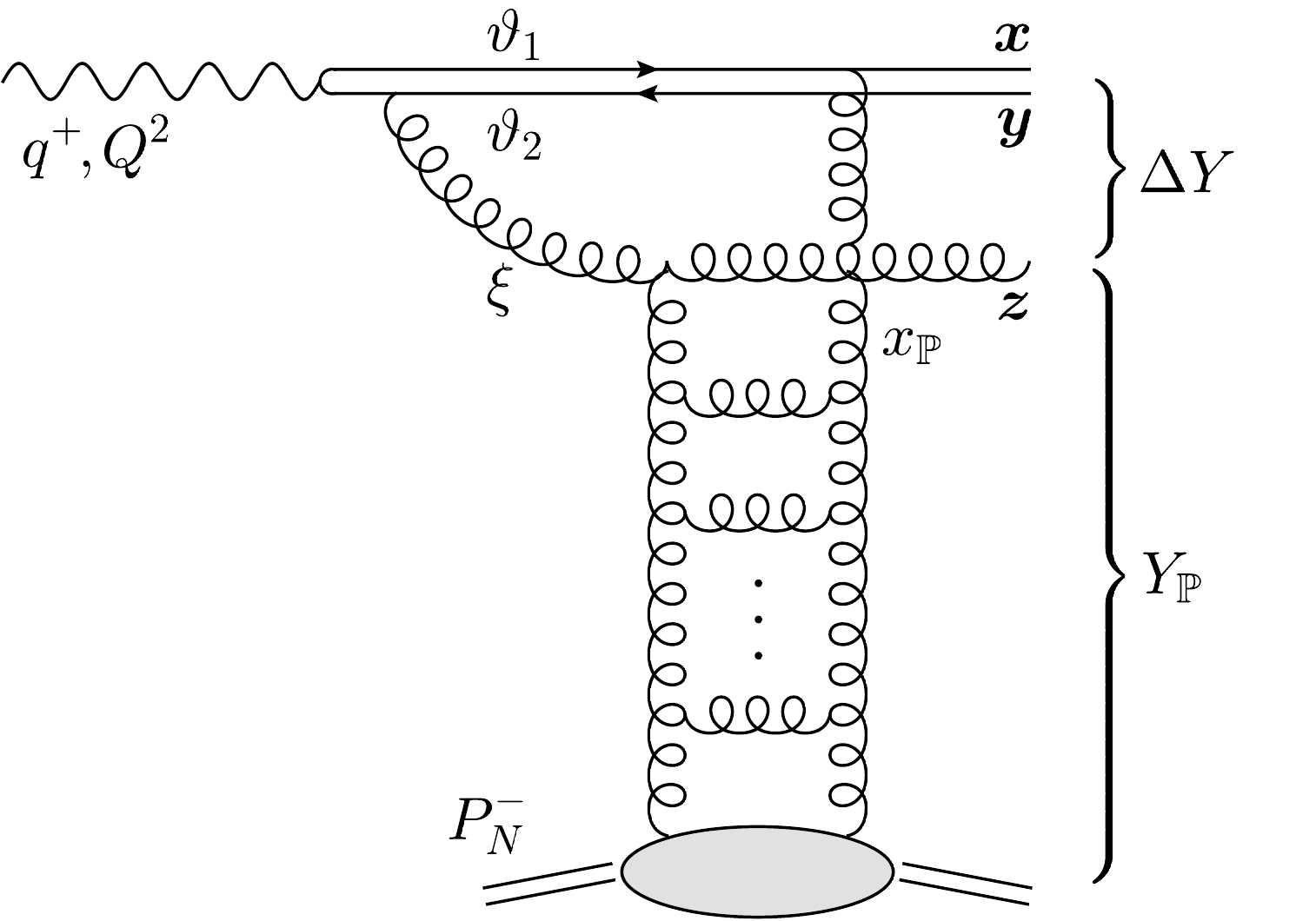}}}
\caption{\small Schematic representation of a Feynman graph contributing to diffractive trijet production. The 
colourless, ``Pomeron'', exchange  is represented by the gluon ladder. }
 \label{fig-diff}
\end{figure}

 The elastic amplitude will be first constructed in transverse coordinate space, to take profit of
the eikonal approximation. We denote the transverse coordinates of
the 3 partons as $\bx,\,\by$ and $\bz$ (see Fig.~\ref{fig-diff}). For the $q\bar q$ pair, we shall also use
$\bm{r}=\bm{x}-\bm{y}$ (the transverse separation)
 and $\bm{b}=\vartheta_1\bm{x}+\vartheta_2\bm{y}$ (the centre-of-energy).

To characterise the rapidity distribution of the final state, we use the
standard variables for diffraction,
\begin{align}\label{betaxP}
\beta\equiv \,\frac{Q^2}{Q^2+M_{q\bar q g}^2},\qquad
x_{\mathbb{P}}\equiv \,\frac{Q^2+M_{q\bar q g}^2}{2q\cdot P_N}\,, 
\end{align}
with the diffractive mass $ M_{q\bar q g}^2 \equiv (k_1+k_2+k_3)^2 $, or
\begin{align}\label{MX}
 \hspace*{-.6cm}
 M_{q\bar q g}^2  = \frac{k^2_{1\perp}}{\vartheta_1} +\frac{k^2_{2\perp}}{\vartheta_2} 
 +\frac{k^2_{3\perp}}{\vartheta_3}=M_{q\bar{q}}^2+K_\perp^2 + \frac{k_{3\perp}^2}{\xi}\,,
\end{align}
where $M_{q\bar{q}}^2 \equiv (k_1+k_2)^2 =P_\perp^2/(\vartheta_1\vartheta_2)$.
Notice that $\beta x_{\mathbb{P}}=\xbj$, or $Y=\Delta Y+Y_{\mathbb{P}}$, where $Y=\ln(1/\xbj)$ 
is the rapidity difference between the target and the virtual photon, $\Delta Y=  \ln(1/\beta)$
is the rapidity phase-space occupied by the $q\bar q g$  system, and $Y_{\mathbb{P}}=\ln(1/x_{\mathbb{P}})$ 
is the rapidity gap between this system and the target.    

 \paragraph{The gluon dipole picture.}
The aforementioned hierarchy of transverse scales implies an appealing physical picture. The 
 hard scale $P_\perp$ controls the size $\br=\bx-\by$ of the $q\bar q$ pair, $r \sim 1/P_\perp$, whereas
the gluon transverse momentum $k_{3\perp}$ controls the transverse separation $\bm{R}= \bm{z}-\bm{b}$
between the gluon and its sources: $R\sim 1/k_{3\perp}$. So, when $k_{3\perp}\ll P_\perp$, we also have $R \gg r$,
and the $q\bar q g$ projectile effectively scatters as a {\it gluon-gluon (gg) dipole}. One leg of this dipole is
the emitted gluon and the other leg is made
with the  $q\bar q$ pair, which remains in a colour octet state after the gluon emission.
A similar picture has  been  used for computing the $q\bar qg$  contribution to the 
diffractive structure function  \cite{Wusthoff:1997fz,GolecBiernat:1999qd}.

The validity of this picture also depends upon the longitudinal scales in the problem. For the typical 
events, the gluon formation time $\tau_3\!=\!2k_3^+/k_{3\perp}^2$ with $k_3^+=\xi q^+$
should not exceed the coherence time $\tau_q\!=\!2q^+/Q^2$ of the virtual photon. This 
implies $\xi \lesssim k_{3\perp}^2/Q^2$. 

In the correlation limit $k_{3\perp}\!\sim \!Q_s\ll Q$, the gluon is {\it soft}, $\xi\lesssim Q_s^2/Q^2\ll 1$, which 
greatly simplifies the evaluation of its emission vertex. 
However, $\xi$ should not be {\it too} small either, since we would like to avoid the emission of additional
soft gluons, which would modify the colour flow and thus spoil the gluon dipole picture. The 
phase-space for soft gluon emissions by the $q\bar q$ pair is $\Delta Y=  \ln(1/\beta)$ and we
shall require $\alpha_s\Delta Y\ll 1$. Using  $M_{q\bar q g}^2\sim {Q_s^2}/{\xi}$ for very small $\xi$,
cf.  \eqn{MX}, we conclude that our picture applies to $\xi$ values
obeying $e^{-1/\alpha_s}\ll \xi Q^2/Q_s^2 \lesssim 1$. This is a parametrically wide range when the coupling is weak.

The most interesting region for studying saturation is the upper end of this window at 
$\xi\sim Q_s^2/Q^2$, since then $\Delta Y\lesssim 1$ and the rapidity gap is as large as possible:
$Y\simeq Y_{\mathbb{P}}$.  Hence most of the total rapidity $Y$ 
is used for the high-energy evolution of the ``Pomeron'' --- the colourless
exchange between the jets and the target, as materialised by the gluon dipole scattering amplitude.
In practice, we shall mostly work with smaller values $\xi \ll  Q_s^2/Q^2$,
since this allows for an important simplification: the gluon emission can be computed in 
the eikonal approximation and then it factorises. 
The generalisation to larger values $\xi\sim Q_s^2/Q^2$ will be briefly discussed towards
the end and further detailed in a subsequent publication \cite{wip}.

 These kinematical constraints are (at least marginally) consistent with the range 
 that should  be covered by the EIC \cite{Accardi:2012qut,Aschenauer:2017jsk}; e.g. for $\xbj\!=\!10^{-3}$,
one will be able to measure  DIS processes with $Q^2\le 10$~GeV$^2$, whereas the nuclear
(Pb) saturation momentum is estimated as $Q_s^2\simeq 1.5$~GeV$^2$. Clearly, the situation
would be even more favorable (higher energies, smaller $\xbj$, and larger $Q^2$)
at the Large Hadron-Electron Collider \cite{LHeC:2020van}.

 \paragraph{The diffractive trijet cross-section.}
 When the gluon is sufficiently soft ($k_{3\perp}\ll P_\perp\sim Q$ and  $\xi \ll  k_{3\perp}^2/Q^2$),
its emission can be computed  in the eikonal approximation. For a virtual photon with transverse polarisation\footnote{The
results for a longitudinal photon are readily obtained by replacing
 $\vartheta_1^2 +\vartheta_2^2 \to 4 \vartheta_1 \vartheta_2$ in
 \eqref{3jetsD} and $P_{\perp}^4 + \bar{Q}^4 \to 2 P_{\perp}^2 \bar{Q}^2$ in \eqref{sumhard}.},
 a homogeneous target with transverse area  $S_\perp$,  and  three flavours of massless quarks, $f=u, d, s$,
 the trijet cross-section reads (with $\alpha_{em}\!=\!e^2/4\pi$)   \cite{wip}
\begin{widetext}
\vspace{-0.5cm}
\begin{align}\label{3jetsD}
\frac{\der \sigma_{\rm D}^{\gamma^*_T A \rightarrow q\bar{q}g A' X}}{\der \vartheta_1 \der \vartheta_2 \der\xi \der^2 \bm{P} 
 \der^2 \bm{K}  \der^2 \bk_3} = S_\perp\frac{ \alpha_{em} 
 N_c}{2\pi^4}\,\Big(\sum e_{f}^{2}\Big)
\big(\vartheta_1^2+\vartheta_2^2\big)\, \delta(1-\vartheta_1-\vartheta_2)\,\delta^{(2)}({\bm{K}+\bm{k}_3})
 \frac{\alpha_s C_F}{\xi}
 \sum_{lj}\big|\mathcal{A}_{q\bar q g}^{lj}\big|^2, \quad
\end{align}
 \vspace{-0.4cm}
\end{widetext}
 where the elastic amplitude takes a factorised form,
 \beq
\mathcal{A}_{q\bar q g}^{lj}=\mathcal{H}^{li}(\bP,\bar Q)\mathcal{G}^{ij}(\bK,Y_{\mathbb{P}}).
\eeq
  The hard  factor (with  $\int_{\bm{r}}\!=\!\int\! \der^2\bm{r}$ and $\bar Q^2\equiv \vartheta_1 \vartheta_2 Q^2$)
\begin{align}   \label{Hij}
\mathcal{H}^{li}(\bP,\bar Q)= \frac{1}{2\pi}\int_{\br} e^{-i \bm{P} \cdot \bm{r}}\, \frac{r^lr^i}{r}
  \, \bar Q \rmK_1(\bar Q r), \end{align}
    describes the decay  $\gamma_T^*\to q\bar q$ of the  virtual photon 
and the vertex $\propto r^i$ for the emission of a transverse gluon from the small $q\bar q$
dipole. The tensorial distribution
  \begin{align}   \label{Gij}
 \hspace*{-.4cm}
 \mathcal{G}^{ij}(\bK, Y)&\equiv \int_{\bm{R}}\,e^{i\bm{K}\cdot \bm{R}}\left(\delta^{ij}-\frac{2{R}^i{R}^j}{{R}^{2}}\right)
\frac{{\mathcal{T}}_{g}({R}, Y)}{2\pi {R}^{2}} \nonumber \\*[0.2cm]
&=\bigg(\frac{K_\perp^i K_\perp^j}{K_\perp^2}- \frac{\delta^{ij}}{2}\bigg)\mathcal{G}(K_\perp, Y),
\end{align}
encodes the spatial distribution of the gluon emission  by the
small $q\bar q$ pair (a dipolar colour field) together
with the scattering between the effective $gg$ dipole and the target.
The scattering amplitude  ${\mathcal{T}}_g({R}, Y)$ is defined as (we recall that 
$\bm{R}= \bm{z}-\bm{b}$)
 \begin{align}
\label{Sdipole}
\mathcal{T}_g(R,Y) = 
1-\frac{1}{N_{c}^2-1}\left\langle\mathrm{tr}\big(U_{\bm{z}}U^{\dagger}_{{\bm{b}}}\big)\right\rangle_Y,
\end{align}
where $U_{\bm{z}},\,U^{\dagger}_{{\bm{b}}}$ are Wilson lines in the adjoint representation and the 
brackets denote the CGC average over the colour fields in the target \cite{Iancu:2003xm,Gelis:2010nm}.
The CGC weight function includes the high-energy, BK/JIMWLK, evolution
\cite{Balitsky:1995ub,Kovchegov:1999yj,JalilianMarian:1997jx,JalilianMarian:1997gr,Kovner:2000pt,Iancu:2000hn,Iancu:2001ad,Ferreiro:2001qy}, 
up to the rapidity scale $Y$.

Using the second line of \eqn{Gij}, one finds
 $ \sum_{lj}|\mathcal{A}_{q\bar q g}^{lj}|^2 =\frac{1}{4} \sum_{li}| \mathcal{H}^{li}|^2\,\mathcal{G}^2$ with
\beq\label{sumhard}
 \sum_{li}\big| \mathcal{H}^{li}\big|^2(\bP,\bar Q)
= 2 \frac{P_{\perp}^4 + \bar{Q}^4}{(P_{\perp}^2 + \bar{Q}^2)^4}\,.\eeq

 \paragraph{The hard dijet cross-section.}
The cross-section for the diffractive production of the $q\bar q$ dijets is obtained from 
 \eqref{3jetsD} by integrating out the kinematical variables $\bk_3$ and $\xi$ of the unmeasured
gluon jet. The integral over $\bk_3$ is trivial and yields $k_{3\perp}=K_\perp$.
The value of $\xi$ is in fact fixed by the rapidity gap: Eqs.~\eqref{betaxP}--\eqref{MX} imply
\beq\label{xixP}
\frac{\rmd\xi}{\xi}
=\frac{\rmd x_{\mathbb{P}}}{x_{\mathbb{P}}-x_{q\bar q}} \quad\mbox{with}\quad
 x_{q\bar q}\equiv \frac{Q^2+M_{q\bar q}^2+K_{\perp}^2}{2P_N\cdot q}\,.
\eeq
When $\xi \ll  K_{\perp}^2/Q^2$, the diffractive mass \eqref{MX} is dominated by the soft gluon,
$M_{q\bar q g}^2\simeq {K_{\perp}^2}/{\xi} \gg M_{q\bar q}^2\sim Q^2$,
hence $x_{q\bar q} \ll  x_{\mathbb{P}}$ and $\rmd\xi/\xi
\simeq {\rmd x_{\mathbb{P}}}/{x_{\mathbb{P}}}=\rmd Y_{\mathbb{P}}$.
The dijet cross-section per unit rapidity gap is then obtained as
\begin{align}\label{3jetsD1}
\hspace*{-.9cm}
 \frac{\der \sigma_{\rm D}^{\gamma^*_T A \rightarrow q\bar{q}A' X}}{\der \vartheta_1 \der \vartheta_2\der^2 \bm{P} \der^2 \bm{K}\rmd Y_{\mathbb{P}}}=H(x_{q\bar q}, Q^2, P_{\perp}^2)
\frac{\der \,xG_{\mathbb{P}}(x, x_{\mathbb{P}}, K_\perp^2)}{\der^2\bm{K}},
 \end{align}
 with the hard impact factor:
 \begin{align}\label{Hard}
\hspace*{-.6cm}
H(x_{q\bar q}, Q^2, P_{\perp}^2)&\equiv
 {\alpha_{em}\alpha_s}\Big(\sum e_{f}^{2}\Big) \,\delta(1-\vartheta_1-\vartheta_2) \nonumber\\
 &\times 
 \left(\vartheta_1^{2}+\vartheta_2^{2}\right)
\frac{P_{\perp}^4 + \bar{Q}^4}{(P_{\perp}^2 + \bar{Q}^2)^4}\,,
 \end{align}
and the unintegrated gluon distribution of the  Pomeron:
\begin{align}\label{dGP}
\frac{\der \,xG_{\mathbb{P}}(x, x_{\mathbb{P}},K_\perp^2)}{\der^2\bm{K}}\equiv S_\perp 
\frac{N_c^2-1}{8\pi^4} \,
 \big[\mathcal{G}(K_\perp, Y_{\mathbb{P}})\big]^2.
 \end{align}
 Here, $x\equiv x_{q\bar q}/x_{\mathbb{P}}\ll 1$ is the fraction of the Pomeron longitudinal momentum transferred to the
hard dijets and  $\mathcal{G}(K_\perp,Y_{\mathbb{P}})$ is related to the gluon dipole amplitude via
\begin{align}\label{Gscalar}\hspace*{-.4cm}
\mathcal{G}(K_\perp,Y_{\mathbb{P}})
 &=2 \int_0^{\infty} \frac{\rmd R}{R}\,
	{\rm J}_2 (K_{\perp} R) \mcal{T}_g(R, Y_{\mathbb{P}}).
\end{align}
This Bessel transform is controlled by dipole sizes $R\lesssim 1/K_\perp$. Let us consider two interesting limits:

\texttt{(i)} for large momenta $K_{\perp}\gg Q_s(Y_{\mathbb{P}})$, we use the single scattering approximation
where $\mathcal{T}_g(R, Y_{\mathbb{P}})$ is proportional to dipole area 
$R^2$ and to the target gluon distribution per unit transverse area;
this gives
  \begin{align}
	\label{hghigh2}\hspace*{-.2cm}
	\mathcal{G}(K_{\perp}, Y_{\mathbb{P}}) \simeq\,\frac{4\pi^2N_c}{N_c^2-1}\,
		\frac{\alpha_s}{ K_{\perp}^2} \,\frac{\der xG(x,K_\perp^2)}{\der^2\bm{b}}\bigg|_{x=
		x_{\mathbb{P}}}.
\end{align}

\texttt{(ii)} for lower momenta $K_{\perp}\ll Q_s(Y_{\mathbb{P}})$, 
we use the black disk limit $\mcal{T}_g=1$ to find
\begin{align}
\label{hglow}
	\mathcal{G}(K_{\perp}, Y_{\mathbb{P}}) \simeq 1.
	\end{align}

In the McLerran-Venugopalan (MV) model  \cite{McLerran:1993ni,McLerran:1994vd},
one can quasi-exactly compute the integral in \eqn{Gscalar} and thus find a global approximation 
interpolating between the two above limits:  
\begin{align}
\label{hgkMV}\hspace*{-.9cm}
\mathcal{G}(K_{\perp})\!
	 =\!\frac{Q_{A}^2}{ K_{\perp}^2} 
	 \ln \frac{K_{\perp}^2}{\Lambda^2}	\!
	 \left[1\!-\exp\left\{- \,\frac{K_\perp^2 }{Q_{A}^2  \ln ({K_{\perp}^2}/{\Lambda^2})}\right\}\right]\!.
\end{align}
The scale $Q_{A}^2  \propto A^{1/3}$ is  related to
the saturation momentum of the MV model via $Q_s^2=Q_{A}^2 \ln(Q_s^2/\Lambda^2)$ \cite{Iancu:2003xm}.

To summarise, the distribution  $\big[\mathcal{G}(K_\perp)\big]^2$ which enters the dijet cross-section
\eqref{3jetsD1} is of order one when $K_\perp\lesssim
Q_s$, but is rapidly decreasing, as $1/K_\perp^4$, when $K_\perp\gg Q_s$. This means that the 
typical value of the dijet momentum imbalance is  $K_\perp\sim Q_s(Y_{\mathbb{P}})$, as anticipated.
So, albeit relatively hard ($P_\perp^2\sim Q^2  \gg Q_s^2$), this diffractive process is
strongly sensitive to gluon saturation.

This sensitivity is stronger than for {\it inclusive} dijets in the correlation limit: 
in that case too, the saturation effects are important when
$K_\perp\lesssim Q_s$. However, at larger momenta $K_\perp\gg Q_s$,
the cross-section decays only as $1/K_\perp^2$. Hence most of the inclusive dijet
events lie in the tail of the distribution at large momenta ($Q_s\ll K_\perp\ll P_\perp$), where saturation is unimportant.

Implicit in Eqs.~\eqref{3jetsD1}--\eqref{dGP} is the shift to a new physical picture, 
where the gluon is emitted by the Pomeron and the small $q\bar q$ dipole inelastically scatters off this gluon,
thus acquiring a momentum imbalance $\bK$. This picture would be manifest if the calculation was performed
in the target light-cone gauge $A^-=0$ (see e.g. \cite{Levin:1992bz}), 
as opposed to the gauge $A^+=0$ implicitly used in deriving 
\eqn{3jetsD}. This {\it a posteriori} change of picture is possible due to the strong hierarchy of transverse
momentum scales, $P_\perp^2\sim Q^2  \gg K_\perp^2\sim Q_s^2$, which is the premise of the collinear factorisation.

Indeed the factorised structure of the diffractive cross-section \eqref{3jetsD1} is
similar to the TMD factorisation for the inclusive dijet production in the correlation
limit \cite{Dominguez:2011wm,Metz:2011wb,Dominguez:2011br}. The hard impact factor is {\it exactly} the same, whereas
the Weiszäcker-Williams gluon TMD is  replaced by the unintegrated gluon distribution of the Pomeron, \eqn{dGP}.

\paragraph{Collinear factorisation and the gluon distribution of the Pomeron.}
When $P_\perp\gg K_\perp$, the dijet cross-section is expected to receives radiative corrections enhanced by 
 the Sudakov  double logarithm $\ln^2(P_\perp^2/K_\perp^2)$ \cite{Mueller:2013wwa}. Such corrections
 could modify the $K_\perp$--distribution of the measured dijets and thus hinder
  the signal of gluon saturation.  To avoid  this problem, we propose to integrate the cross-section over $K_\perp$, 
  up to the hard scale $P_\perp$:
\begin{align}\label{gluondipColl}
 \hspace*{-.8cm}
\frac{\rmd\sigma_D^{\gamma_T^* A\rightarrow q\bar q A' X}}{\der \vartheta_1 \der \vartheta_2\der^2 \bm{P} \rmd Y_{\mathbb{P}}} =
H(x_{q\bar q}, Q^2, P_{\perp}^2)\, xG_{\mathbb{P}}(x, x_{\mathbb{P}}, P_\perp^2),
 \end{align}
 where  $x= x_{q\bar q}/x_{\mathbb{P}}$ and $xG_{\mathbb{P}}(x, x_{\mathbb{P}}, P_\perp^2)$ is the
 gluon distribution of the Pomeron, as obtained by integrating \eqn{dGP} over $K_\perp$. \eqn{gluondipColl}
is recognised as the collinear factorisation for the diffractive process at hand.
 
 Unlike for the {\it inclusive} dijets, where the integration over $K_\perp$ has the drawback
to wash out the sensitivity to gluon saturation, there is no similar difficulty  for the diffractive dijets:
the function $\big[\mathcal{G}(K_\perp)\big]^2$ is rapidly decreasing when $K_\perp\gg Q_s$, so
its integral is dominated by $K_\perp\sim Q_s$ and is independent of the upper cutoff $P_\perp$:
   \beq\label{G2int} \hspace*{-.4cm}
\int\rmd^2\bK\, \big[ \mathcal{G}(K_\perp)\big]^2
=4\int_{\bm{{R}}}\frac{\mathcal{T}_{g}^2({R})}{{R}^{4}}
\simeq \pi\kappa\, Q_{s}^2(Y_{\mathbb{P}})\,,
\eeq
with $\kappa$ a number depending on our approximation for
 $\mathcal{T}_{g}({R})$; e.g., the MV model \eqref{hgkMV} yields $\kappa =2\ln 2$.
 Thus, quite remarkably, 
 the physics of saturation determines  the Pomeron gluon distribution which enters 
  the collinear factorisation at the scale $Q^2\gtrsim Q_s^2$.
 In turn, this can be used as an initial condition for the DGLAP evolution towards larger values $Q^2\gg Q_s^2$.

\paragraph{Initial condition for DGLAP evolution from gluon saturation.}
So far, the Pomeron UGD  in \eqn{dGP} has been constructed for  $x\ll 1$,
but the initial condition for the DGLAP  equation is needed for generic values $x\le 1$.
Using $(1-x)/x\sim {K_\perp^2}/{\xi Q^2}$ (cf. Eqs.~\eqref{betaxP}, \eqref{MX} and \eqref{xixP}), one sees
that $x\sim\order{1}$ corresponds to $\xi\gtrsim K_\perp^2/Q^2$, i.e. to gluon emissions
with relatively large formation times $\tau_3\gtrsim\tau_q$,  for which the eikonal 
approximation does not apply anymore. So long as we stay in the correlation limit $Q^2\gg  K_\perp^2$,
one can still use the condition $\xi\ll 1$ to simplify the gluon emission vertex. However,
one now needs a more accurate treatment of the light-cone energy denominators.
This will be detailed in  Ref.~\cite{wip}, but the main results are quite simple: to parametric accuracy,
it suffices to restrict the integral in \eqn{Gscalar}  to dipole sizes
\beq\label{Rmax}
R^2\,\lesssim\,\frac{1}{\xi Q^2}\,\sim\,\frac{1-x}{x K_\perp^2}\,.
 \eeq
 Hence, for  $x\sim\order{1}$, the transverse separation $R$ between the gluon and the $q\bar q$ pair
at the time of scattering is considerably smaller than its final value $R_f\sim 1/K_\perp$ at the time of emission.
This is understood as follows: in a quantum emission, the gluon separates from its sources 
via diffusion, $R^2(\tau)\sim \tau/k_3^+$, until a time $\tau\sim\tau_3$, when $R(\tau_3)=R_f$ and the gluon
is freed. However, when $\tau_3$ is larger than the typical scattering time (here, of order
$\tau_q$), then $R^2(\tau_q)\sim\tau_q/k_3^+\sim 1/(\xi Q^2)$
is necessarily smaller than $R_f^2$, by a factor  $\tau_q/\tau_3=(1-x)/x$.

\begin{figure}[t] \centerline{
\centerline{\includegraphics[width=0.98\columnwidth]{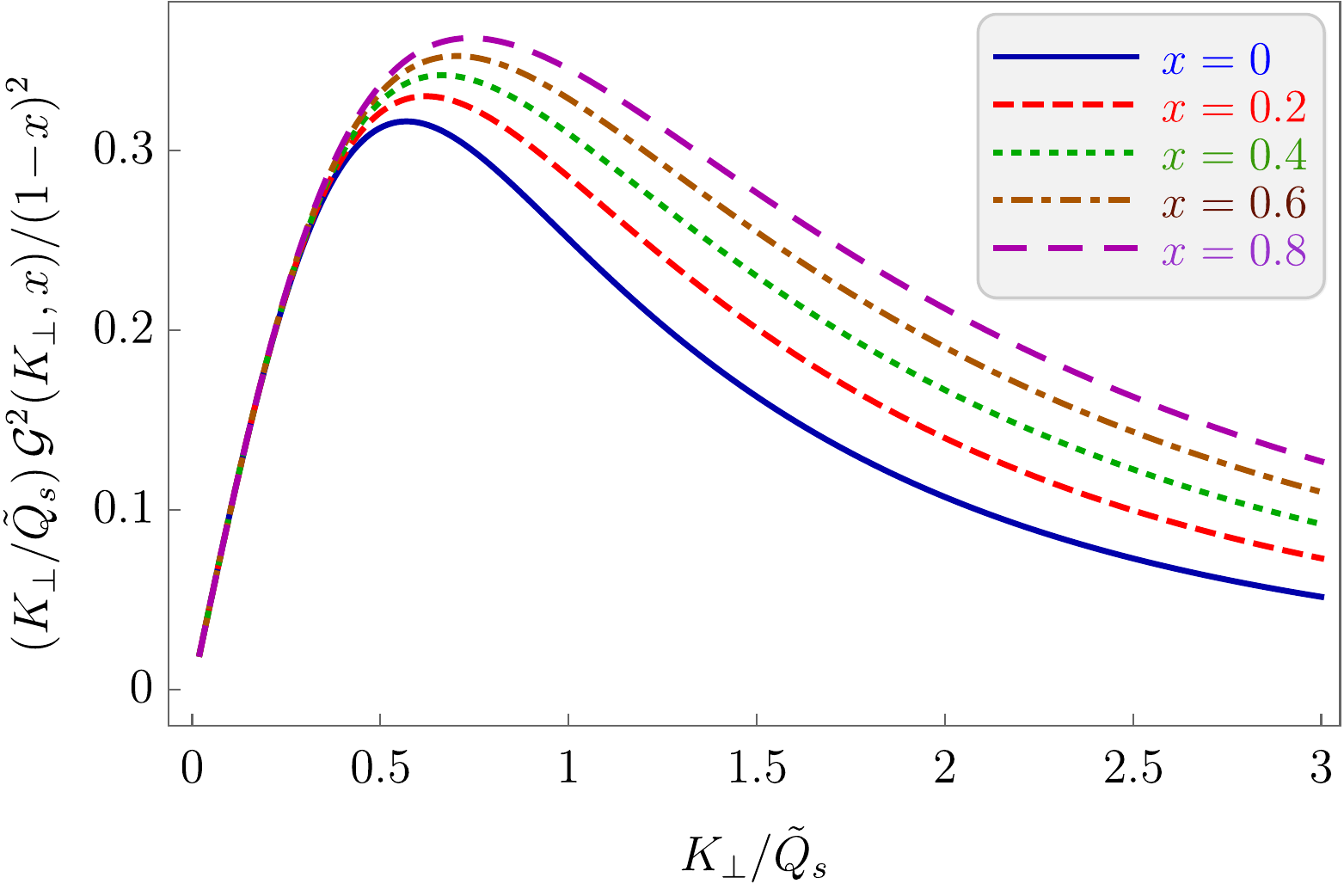}}}
\caption{\small The UGD of the Pomeron $\mathcal{G}(K_\perp,x)$ obtained from an extension
of \eqn{Gscalar} valid for generic values of $x\equiv x_{q\bar q}/x_{\mathbb{P}}$ \cite{wip}. 
We  use the MV model for $\mathcal{T}_{g}({R})$ and plot the function $(K_\perp/\tilde Q_s)[\mathcal{G}(K_\perp,x)/(1-x)]^2$,
with  $\tilde Q_{s}^2= (1-x)Q_{s}^2$, as a function of  $K_\perp/\tilde Q_s$,
for various values of $x$.}
 \label{fig-GxK}
\end{figure}

The main effect of this new constraint  is to introduce an effective,
 $x$--dependent, saturation momentum, $\tilde Q_{s}^2(x,Y_{\mathbb{P}})\equiv (1-x)Q_{s}^2(Y_{\mathbb{P}})$,
 which separates between weak and strong scattering  for generic $x$ \cite{wip}. Roughly speaking, the function
 $\mathcal{G}(K_\perp,x)$  can be obtained from its value at small-$x$   by replacing $Q_s\to \tilde Q_{s}(x)$
  and multiplying the result by $(1-x)$ (see Fig.~\ref{fig-GxK}).  
 Integrating over $K_\perp^2$ gives the following 
 estimate for the Pomeron gluon distribution at $Q^2\gtrsim Q_s^2(Y_{\mathbb{P}})$
 and arbitrary $x\equiv x_{q\bar q}/x_{\mathbb{P}}$ :
   \begin{align}\label{ICDGLAP}
\hspace*{-.6cm}
xG_{\mathbb{P}}(x, x_{\mathbb{P}}, Q^2)=S_\perp (1-x)^2
\frac{N_c^2-1}{(2\pi)^3}\kappa\,Q_s^2(Y_{\mathbb{P}}).
\end{align}
This expression, emerging from first principles, can be used as an initial condition for the DGLAP equation.

\paragraph{Summary and perspectives.} We have shown that hard diffractive production of 3 jets in the correlation limit
is a sensitive probe of gluon saturation in DIS at small $x$. This sensitivity persists after integrating over the
kinematics of the soft jet to obtain the collinear factorisation for the hard dijets.
Our analysis has focused on the main phenomena. 
Explicit results have been presented in the MV model, for simplicity. The effects of the pQCD
evolution can be numerically important and also conceptually interesting, because of
the interplay between the BK/JIMWLK and the DGLAP evolutions.
We will include these effects in a subsequent study \cite{wip}. It would be interesting to
better understand the interplay between the Sudakov effects and gluon saturation
 on the dijet distribution in the momentum imbalance $K_\perp$. Last but not least, one may
think about a similar process in ultraperipheral proton-nucleus, or nucleus-nucleus, collisions at the LHC, 
where the available energies are much higher.

\smallskip
\acknowledgements{
\noindent{\bf Acknowledgements} 
The work of E.I. is supported in part by the Agence Nationale de la Recherche project 
 ANR-16-CE31-0019-01.   The work of A.H.M.
is supported in part by the U.S. Department of Energy Grant \# DE-FG02-92ER40699. 
}

\providecommand{\href}[2]{#2}\begingroup\raggedright\endgroup

\end{document}